\documentclass[referee]{raa}            
\usepackage{graphicx,times}             
\usepackage{longtable}
\usepackage{rotating}

\begin{document}
   \title{Noise reduction methods in analysis of near infrared lunar occultation light curves for high angular resolution measurements}

   \volnopage{Vol.0 (200x) No.0, 000--000}      
   \setcounter{page}{1}          

   \author{Tapas Baug
      \inst{1}
   \and Thyagarajan Chandrasekhar
      \inst{1}
    }

\institute{Physical Research Laboratory, Ahmedabad-380009, India;\it tapasb@prl.res.in}
\date{Received~~2013 May 01; Accepted~~2013 May 01}

\abstract{Lunar occultation (LO) technique in the near-infrared provides angular resolution down to milliarcseconds on the occulted source even with ground-based 1m class telescopes. LO observations are limited to brighter objects because they require high signal to noise ratio (S/N $\sim$40) for proper extraction of angular diameter values. Hence, methods to improve the S/N ratio by reducing noise using Fourier and Wavelet transforms have been explored in this study. A sample of 54 near-infrared LO light curves observed with IR Camera at Mt Abu observatory has been used. It is seen that both Fourier and Wavelet methods have shown improvement in S/N, compared to the original data. However, the application of wavelet transforms results in slight smoothening of the fringes resulting in a higher angular diameter value. Fourier transforms which reduce discrete noise frequencies do not distort the fringe. Fourier transform method seems to be effective in S/N improvement, as well as improved model fit particularly 
in the fainter regime of our sample. They also provide a better model fit for brighter sources in some cases though there may not be significant S/N improvement. 
}

   \authorrunning{T. Baug \& T. Chandrasekhar}            
   \titlerunning{Noise reduction in NIR lunar occultation light curves }  

   \maketitle

%
%
\section{Introduction}           
\label{sect:intro}
The lunar occultation (LO) technique in the near infrared (NIR) is capable of providing high angular resolution information down to 1 milli-arcsecond on occulted stellar sources. It is among the most rapid of celestial events with the entire event getting over in just a few hundred milliseconds. The acquisition of LO fringes requires high speed photometry of sampling in the milliseconds range. The LO fringes are due to Fresnel diffraction pattern produced by the sharp edge of the lunar limb, of the light from the occulted stellar source. Various effects like the optical bandwidth of filter used, the averaging effect of the telescope aperture, subtraction of sky background, the time response of the detector system have to be correctly accounted for, in oder to extract proper angular diameter value (in milliarcseconds) of the occulted source. One of the main drawbacks in analysis is the limitation imposed by the S/N of the detected diffraction pattern. A minimal signal to noise ratio (S/N $\sim$ 40) is usually required to have a good angular diameter measurement. An LO light curve with very high S/N ($>$ 100) in the fringe region can also be investigated for signatures of circumstellar material surrounding the star. Also detecting finer structure information of the inner dust shell of sources, for example WR 104, IRC + 10216 (Mondal \& Chandrasekhar, 2002; Chandrasekhar \& Mondal 2001) is only possible with higher S/N light curves. However, the fast sampling and good S/N are conflicting requirements and set a limit on the magnitude of observable sources using the LO technique. With the present detector system (NICMOS IR Camera) sources upto $m_K \sim 5.0$ are observable under good sky conditions. The earlier system (a single element InSb photometer) could reach only upto $m_K \sim 3.0$ but with good sampling down to 1 millisecond. In this paper attempts has been made to explore different noise detection and reduction techniques to reduce the noise content in the LO light curves and thereby enhancing S/N.

\begin{figure}
\centering
  \includegraphics[width=14cm,height= 8cm]{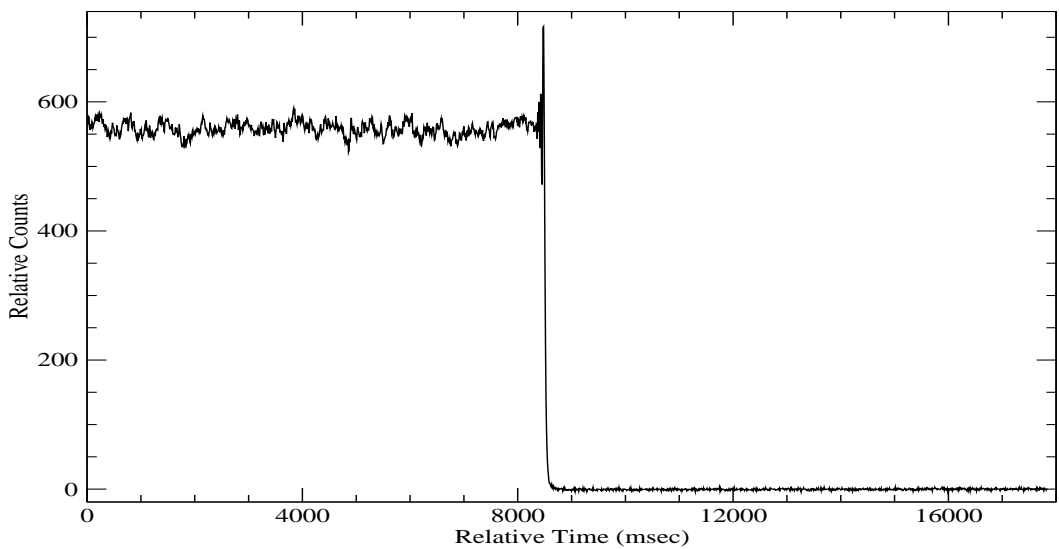}
  \caption[Noise pattern in the LO light curve of bright sources SAO 118044.]{The light curve of one of the brightest sources, SAO 118044 ($m_K = 0.49$), in our sample shows a different noise pattern before and after the occultation. The first part of the light curve (before event) is dominated by source photon noise and low frequency scintillation noise. But the second part of the light curve (after event) is purely dominated by background noise.}
\label{sao118044_time}
\end{figure}

\begin{figure}
\centering
  \includegraphics[width=14cm,height= 8cm]{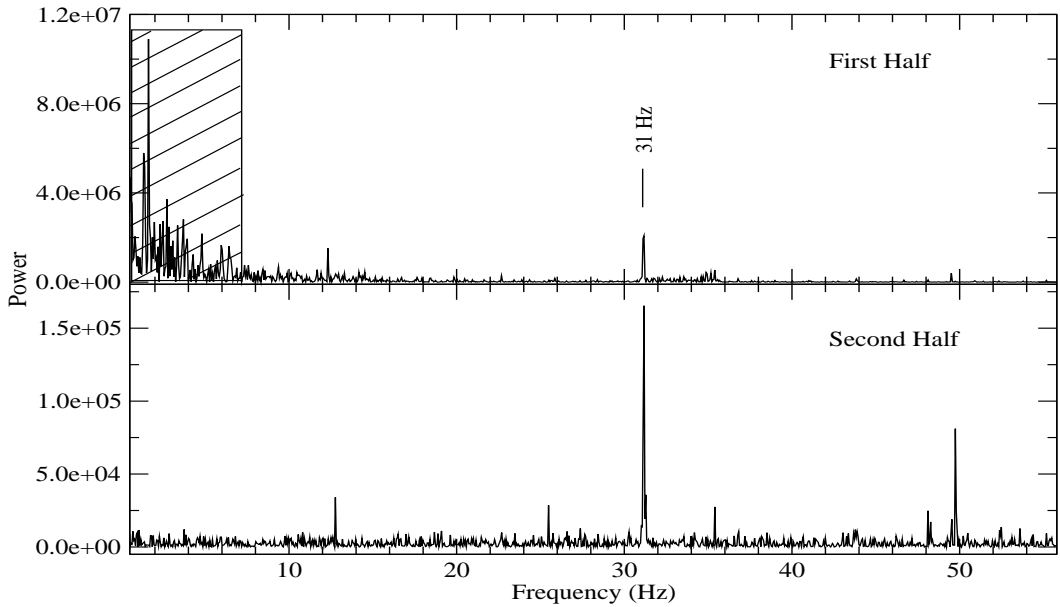}
  \caption[Fourier power spectrum of LO light curve of SAO 118044.]{The power spectrum of the LO light curve of SAO 118044 shown in Fig.\ref{sao118044_time}. Note that different scales has been used in Y-axis in the two halves. The upper panel shows the power spectrum of the first half of the light curve (before occultation) and the lower panel shows the power spectrum of the second half. First half of the light curve is dominated by scintillation and source photon noise. There are many low frequency components as seen inside the shaded box. In the second half these components are absent. The noise at several discrete frequencies like 31 Hz, 50 Hz with sub-harmonics are present in both halves of the light curve.}
\label{sao118044_ft}
\end{figure}

Lunar occultation observations are carried out in near infrared (NIR) K-band (2.2 $\mu m$) to reduce lunar scattered back-ground. Lunar back-ground is higher in other two NIR bands, namely J (1.25 $\mu m$) and H (1.65 $\mu m$) as well as visible bands compared to K-band. Also Lunar occultation in the K-band is capable of providing high angular resolution information of the true photosphere as well as circumstellar shell, if present. It is difficult to acquire shell informations in the other NIR bands J \& H . In this study the late-type giants are mainly studied and they are brighter in the K-band compared to other optical bands. Lunar occultation light curve in the near infrared (NIR; 2.2 $\mu m$) suffers from noise contributions due to both source and back-ground. Scintillation noise is more prominent in case of brighter sources. Scintillation noise is also altitude dependent, lower the altitude of the source more prominent is the scintillation noise. Hence, the recorded light curves show a low frequency variation in the signal level when the star is available. In Fig.\ref{sao118044_time}, the LO light curve of SAO 118044 ($m_K \sim$0.5), low frequency variation of the signal level can be noticed before occultation. Another definite source of noise in the astronomical observations of incoherent sources is source photon noise, which arises due to intrinsic natural variation of the emitted photon flux. It creates high frequency fluctuation in signal level of the observed LO light curve. It can be noticed in the first half of the light curve, high frequency fluctuations created due to stellar photon noise are modulated by low frequency scintillation noise. While the second half of the light curve is mainly dominated by back-ground sky photon noise as the source is absent there. The corresponding power spectra of the light curve SAO 118044 is shown in Fig.\ref{sao118044_ft}. First part of the light curve is dominated by low frequency noise, mainly scintillation noise. While second half of the light curve is dominated by specific noise frequencies.

The S/N of NIR observations is usually limited by sky noise due to scattered moonlight (variable with lunar phase), airglow intensity (variable) and also thermal contribution from telescope optics especially when broad K-band filter is used. From our observed light curves we found that a change over in the dominating noise from source to back-ground noise takes place at K-magnitude of about 3.3 under good sky condition and good altitude of observation ($>$40$^o$). Another significant contribution of noise in the light curves comes from electrical line pick-up (50 Hz) and an 31 Hz noise frequency from the detector electronics. Though it was ascertained to originate from the detector electronics, but attempts to eliminate it in the hardware were not successful. 

Fourier transform (FT) is capable of detecting independent, non-evolving periodicities in the stationary signals. FT have been extensively used in different aspects in astrophysics particularly for identifying and processing stationary time series data. Application of Fourier analysis to identify low frequency correlated noise on binary pulsar data can be found in Kopeikin \& Patapov (2004). In a recent paper Kashlinsky et al. (2012) have used Fourier transforms for measurements of cosmic infrared back-ground fluctuations using SPITZER data. The power spectra of source subtracted fields were examined to eliminate or identify different noise components.

Recently wavelet transform is being widely used for analyzing oscillations, periodicities, as well as sudden change in the observed signal. Decomposition of time series data into signal like and noise like can be found in Polygiannakis et al., 2003. They applied it on sunspot index to examine other periodicities in the solar cycle than  11 years. In a recent paper, Adamakis et al. (2011) used wavelet transform to analyze long term optical light curve of about 76 years of nova RS Ophiuchi. From the wavelet power, they have identified a pre-outburst signature up to 450 days prior. This indicates to have a potential to put an early warning of future outbursts. Wavelet has also been used earlier in analysis of lunar occultation data. In 2008, Fors et al. used wavelet transform to compute the initial guess of the model parameters while they were handling a large data set.

In this paper we have carried out methods to improve S/N of the observed LO light curves by reducing noise content from the observed LO light curves. The noise reduction methods are applied on a sample of 54 light curves upto $m_K \sim 5.0$ obtained using the NICMOS IR camera attached to Mt. Abu telescope. Observational details are given in Table.\ref{src_list}. The sources in the table are arranged in increasing order of RA. First column of the table is the designation of the sources, while the second and third columns are date of observations and J2000.0 coordinates respectively. Fourth column represents the NIR K-magnitude of the source as given in 2MASS catalog (Cutri et al., 2003). Fifth column represents the spectral type of the source collected from the SIMBAD database. Finally, sixth and seventh column respectively correspond to the elevation of the source above the horizon and lunar phase (days after new moon) during the observations.
\begin{center}
\begin{longtable}{lccclcr}
\caption{Details of observations of $Lunar$ $Occultation$ events at Mt. Abu Observatory, observed in near infrared K-band (2.2 $\mu m$/ Band-width 0.4 $\mu m$).}
\label{src_list}\\
\hline
  Source    & Date of      &   Coordinates & $m_K$ & Sp. type &     El     & Lunar \\
            & Observations &    (J2000.0)  &(2MASS)&          & ($^\circ$) & Phase \\
\endfirsthead

\multicolumn{7}{c}{{\tablename\ \thetable{} -- continued from previous page}}              \\
\hline
  Source    & Date of      &   Coordinates & $m_K$ & Sp. type &     El     & Lunar \\
            & Observations &    (J2000.0)  &(2MASS)&          & ($^\circ$) & Phase \\
\hline
\endhead

\hline
\endfoot

\hline
SAO 109252	&      2011 Jan 11    &       00      31      37.360  +09     09      40.26   &       3.74    &       K2	      &       57.8    &       7.22    \\
SAO 92697	&      2009 Nov 29    &       01      55      31.157  +17     11      21.72   &       2.03    &       M6	      &       70.9    &       12.83   \\
SAO 92755$^a$	&      2009 Nov 29    &       02      01      30.386  +17     57      13.79   &       2.50    &       M...	      &       47.0    &       13.02   \\
SAO 92755$^b$	&      2010 Jan 23    &       02      01      30.386  +17     57      13.79   &       2.50    &       M...	      &       82.9    &       8.25    \\
IRC +20037	&      2010 Jan 23    &       02      03      42.621  +18     15      11.69   &       2.90    &       K4III	      &       61.4    &       8.33    \\
UY Ari		&      2011 Jan 13    &       02      05      49.064  +17     29      34.53   &       3.69    &       M5	      &       80.4    &       9.22    \\
AU Ari		&      2011 Jan 13    &       02      08      56.711  +17     34      45.95   &       3.68    &       M0	      &       52.0    &       9.31    \\
SAO 75669	&      2011 Jan 14    &       02      58      48.237  +20     35      39.75   &       3.35    &       M0	      &       78.1    &       10.26   \\
UZ Ari$^a$	&      2009 Nov 30    &       03      01      34.722  +21     48      12.19   &       1.25    &       M8	      &       40.8    &       14.08   \\
UZ Ari$^b$	&      2010 Jan 24    &       03      01      34.722  +21     48      12.19   &       1.25    &       M8	      &       46.4    &       9.41    \\
SAO 75706	&      2011 Jan 14    &       03      03      03.525  +21     10      23.84   &       3.71    &       M0	      &       34.0    &       10.40   \\
IRAS 03333+2359	&      2009 Jan 07    &       03      36      22.250  +24     09      27.75   &       5.07    &       --	      &       55.9    &       10.03   \\
IRC +20066	&      2011 Jan 15    &       03      53      54.698  +23     07      47.36   &       2.60    &       M...	      &       87.3    &       11.27   \\
NSV 1406	&      2011 Jan 15    &       03      54      43.086  +23     19      12.13   &       2.93    &       M4	      &       73.9    &       11.31   \\
SAO 76350	&      2010 Jan 25    &       03      57      26.391  +24     27      43.05   &       3.20    &       K0	      &       85.2    &       10.33   \\
V1134 Tau	&      2009 Mar 03    &       04      04      11.955  +25     23      56.61   &       3.84    &       M0	      &       43.3    &       6.60    \\
SAO 76450	&      2010 Jan 25    &       04      06      04.684  +24     43      55.51   &       3.06    &       S3/3	      &       21.2    &       10.54   \\
IRAS 04320+2519$^a$	&   2009 Dec 29    &       04      35      05.617  +25     26      00.46   &       4.30    &       --	      &       46.0    &       13.35   \\
IRAS 04320+2519$^b$	&   2010 Feb 22    &       04      35      05.617  +25     26      00.46   &       4.30    &       --	      &       78.5    &       8.48    \\
C* 3246		&      2010 Jan 26    &       04      57      20.878  +25     37      44.35   &       3.58    &       C 	      &       60.0    &       11.26   \\
IRAS 05013+2704	&      2009 Mar 04    &       05      04      30.698  +27     08      26.29   &       4.12    &       --	      &       74.1    &       7.54    \\
SAO 76952	&      2011 Mar 12    &       05      05      12.039  +23     47      56.26   &       4.70    &       K0	      &       44.3    &       7.81    \\
SAO 76965	&      2009 Mar 04    &       05      07      05.162  +26     59      45.30   &       4.40    &       F5V	      &       57.8    &       7.59    \\
IRAS 05212+2655	&      2009 Feb 05    &       05      24      19.176  +26     58      22.04   &       3.74    &       --	      &       75.9    &       10.27   \\
IRAS 05232+2400	&      2011 Feb 13    &       05      26      16.887  +24     02      58.82   &       3.94    &       --	      &       72.2    &       10.58   \\
SAO 77229	&      2009 Feb 05    &       05      31      56.397  +26     41      05.94   &       3.87    &       K2	      &       44.0    &       10.46   \\
SAO 77271	&      2009 Dec 30    &       05      34      51.086  +25     53      03.80   &       3.27    &       K7	      &       75.6    &       14.21   \\
SAO 77357	&      2010 Feb 23    &       05      39      32.010  +25     18      45.07   &       4.39    &       M0	      &       69.8    &       9.55    \\
SAO 77474	&      2010 Feb 23    &       05      44      07.749  +25     22      25.59   &       4.14    &       K5	      &       38.6    &       9.65    \\
HD 249571	&      2011 Mar 13    &       05      57      38.274  +23     33      28.98   &       4.50    &       M0III	      &       84.7    &       8.69    \\
SAO 77792	&      2011 Jan 17    &       05      58      38.163  +23     39      57.94   &       2.97    &       M2	      &       58.5    &       13.44   \\
IRAS 05551+2305	&      2011 Mar 13    &       05      58      10.549  +23     06      20.76   &       4.18    &       --	      &       88.0    &       8.71    \\
BI Gem		&      2010 Jan 27    &       06      05      49.697  +25     14      41.56   &       3.17    &       M6	      &       71.9    &       12.34   \\
IRC +30140	&      2009 Mar 05    &       06      09      20.095  +26     31      42.25   &       2.70    &       M1	      &       88.1    &       8.54    \\
TU Gem		&      2009 Jan 09    &       06      10      53.107  +26     00      53.32   &       0.82    &       C 	      &       18.5    &       12.46   \\
BD+26 1131	&      2009 Mar 05    &       06      12      22.576  +26     20      03.86   &       4.06    &       M3IIIe...       &       65.1    &       8.61    \\
IRAS 06165+2431	&      2010 Mar 23    &       06      19      37.503  +24     30      25.59   &       4.87    &       --	      &       80.1    &       7.71    \\
IRAS 06395+2409	&      2010 Feb 24    &       06      42      35.253  +24     06      41.56   &       3.66    &       --	      &       87.7    &       10.54   \\
IRC +20177	&      2010 Jan 28    &       07      20      07.367  +21     58      56.12   &       2.56    &       F0IV	      &       51.2    &       13.57   \\
IRC +20186	&      2009 Feb 07    &       07      40      58.525  +23     01      06.75   &       2.18    &       K5III	      &       84.4    &       12.39   \\
VV Cnc		&      2010 Jan 29    &       08      11      16.298  +19     08      54.59   &       0.94    &       M5	      &       20.7    &       14.27   \\
CW Cnc$^{\dag}$	&      2010 Apr 22    &       09      08      26.526  +13     13      13.64   &       0.11    &       M6	      &       18.3    &       8.28    \\
6 Leo		&      2011 Apr 13    &       09      31      57.591  +09     42      57.04   &       1.98    &       K2.5IIIb        &       42.9    &       10.16   \\
IRC +10210	&      2010 May 20    &       09      41      09.022  +09     53      32.11   &       2.50    &       A5V+..	      &       64.5    &       6.56    \\
SAO 118044	&      2010 Apr 23    &       10      00      12.812  +08     02      39.13   &       0.49    &       M2III	      &       45.3    &       9.23    \\
IRC +00202	&      2009 May 31    &       11      06      02.226  +01     12      38.34   &       2.59    &       M...	      &       38.5    &       7.19    \\
SAO 157613	&      2009 May 06    &       12      56      43.052  -11     56      37.14   &       3.24    &       K5	      &       52.8    &       11.59   \\
CW Oph		&      2011 May 18    &       16      55      39.099  -23     52      25.91   &       2.65    &       M7:	      &       37.5    &       15.50   \\
RT Cap		&      2008 Dec 02    &       20      17      06.516  -21     19      04.30   &       0.31    &       C 	      &       42.3    &       4.70    \\
RS Cap$^{\dag}$	&      2008 Dec 03    &       21      07      15.428  -16     25      21.50   &      -0.22    &       M6-7III	      &       45.8    &       5.82    \\
IRC -10564	&      2008 Nov 06    &       21      30      37.105  -14     10      18.85   &       2.22    &       M4III:	      &       16.7    &       8.78    \\
SAO 145698	&      2010 Dec 11    &       21      50      35.452  -08     58      57.26   &       3.80    &       K0	      &       52.7    &       5.80    \\
SAO 145992	&      2008 Dec 07    &       22      16      52.574  -09     02      24.45   &       3.22    &       K3III:	      &       35.7    &       9.75    \\
TX Psc$^{\dag}$	&      2011 Nov 06    &       23      46      23.520  +03     29      12.15   &      -0.51    &       C 	      &       68.9    &       10.83   \\

\end{longtable}
\small{$^{\dag}$ sources were observed in Narrow CO-band (2.37$\mu m$; BW 0.1 $\mu m$).}
\end{center}


\section{Observations}
A total of 54 occultation light curves have been used in the analysis. All the light curves are observed in the near-infrared broad-$K$-band (2.2 $\mu m$/Band-width 0.4 $\mu m$), except three. Three bright sources are observed in narrow $CO$-band (2.37 $\mu m$/BW 0.1 $\mu m$) to avoid saturation effect and they are marked as $\dag$ in Table.\ref{src_list}. All the LO light curves are observed using the NICMOS IR camera attached to the 1.2 m telescope at Mt Abu Observatory (Lat: 24$^\circ$ 39$^\prime$ 10$^{\prime\prime}$, Long: 72$^\circ$ 46$^\prime$ 47$^{\prime\prime}$ E, Alt: 1680m ), India. The NICMOS camera was used in the imaging sub-array mode of size 10X10 pixels $(5^{\prime\prime} X 5^{\prime\prime}$ on the sky ) to sample the events as rapidly as possible. The sub-array operational mode of the NICMOS camera was first used for LO observations towards the Galactic centre on 2007 September 19 (Chandrasekhar \& Baug 2010). Typically an observing run in the sub-array mode consists of initiating the data acquisition procedure for recording 4800 sub frames about 20 seconds before the predicted time of the event. About 15 seconds after the predicted event time the telescope is rapidly switched to nearby sky to record sky frames. Star becomes ``visible'' only in sky subtracted frames due to the large back-ground level in K-band. Sky subtracted sub-frames are used to derive the light curve. All but one source in this sample were observed in 10X10 pixels. With an integration and a reset time of 3 msec each, it provides a sampling time of 8.926 msec. The source IRC +20177 was observed in 20X20 pixels, to accommodate its visual binary component in the sub-array. It has a sampling time of $~$14 msec.

\section{Analysis}
\subsection{Fourier Denoising method}
Any stationary signal is decomposed into its component frequencies using Fourier Transforms (FT). It converts the signal from time-amplitude to frequency-power domain. But one of the draw-backs of FT is that the time information of the signal is lost in the frequency domain; which means, the specific time of occurrence of any particular frequency is completely lost. In the present case most of the light curves are sampled with 8.926 milliseconds and the corresponding $Nyquist$ frequency is 56.0 Hz. In our LO light curves specific well identified noise frequencies like 31 Hz, 50 Hz and their sub-harmonics are present. Hence, it is useful to apply Fourier transforms to remove these frequencies. The Fourier denoising ($FD$) method is applied to our sample consisting of 54 observed light curves (Table.\ref{src_list}). A {\sc MATLAB} script written by us has been used in the analysis. A standard set of 4096 data points for each light curves have been used for all the transformations.
\begin{figure}
  \begin{minipage}[t]{0.495\linewidth}
  \centering
\includegraphics[width=60mm,height=35mm]{ms1495fig3.eps}
\caption{A step function (inset) and its power spectrum. The power-spectrum shows rapidly decreasing power towards higher frequencies.}
\label{ft_step_fn}
  \end{minipage}
  \begin{minipage}[t]{0.495\textwidth}
  \centering
\includegraphics[width=60mm,height=35mm]{ms1495fig4.eps}
\caption{A step function with slope (inset) and its power spectrum. The power-spectrum shows an oscillatory behavior along with the rapidly decreasing power.}
 \label{ft_step10}
  \end{minipage}
\end{figure}

\begin{figure}[ht]
 \centering
\includegraphics[width=14cm,height= 9cm]{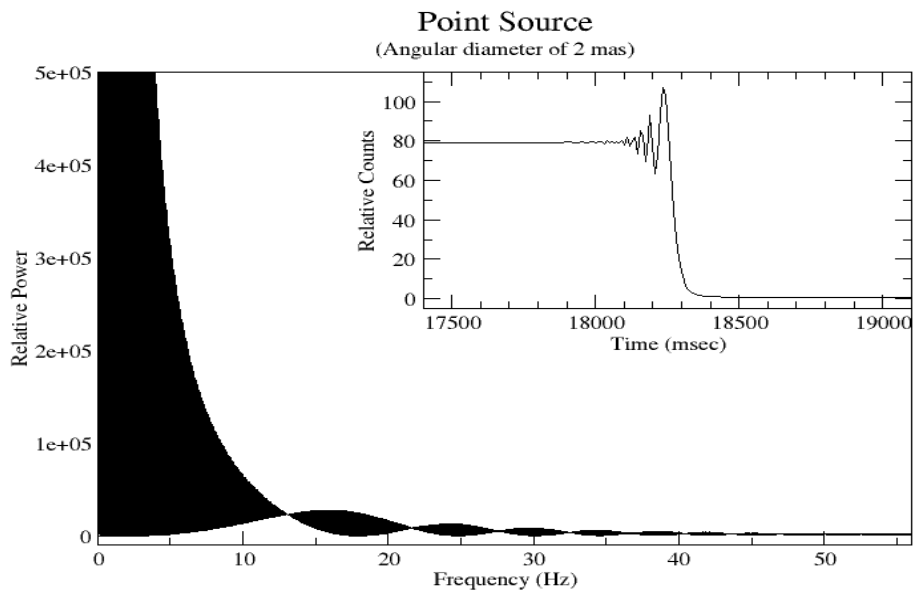}
\caption{The power spectrum of model lunar occultation light curve of a 2 milliarcsec source (inset) is similar to that of the step function with slope, shown in Fig.\ref{ft_step10}.}
\label{ft_of_model}
  \end{figure}

\begin{figure}[ht]
\centering
\includegraphics[width=14cm,height= 9cm]{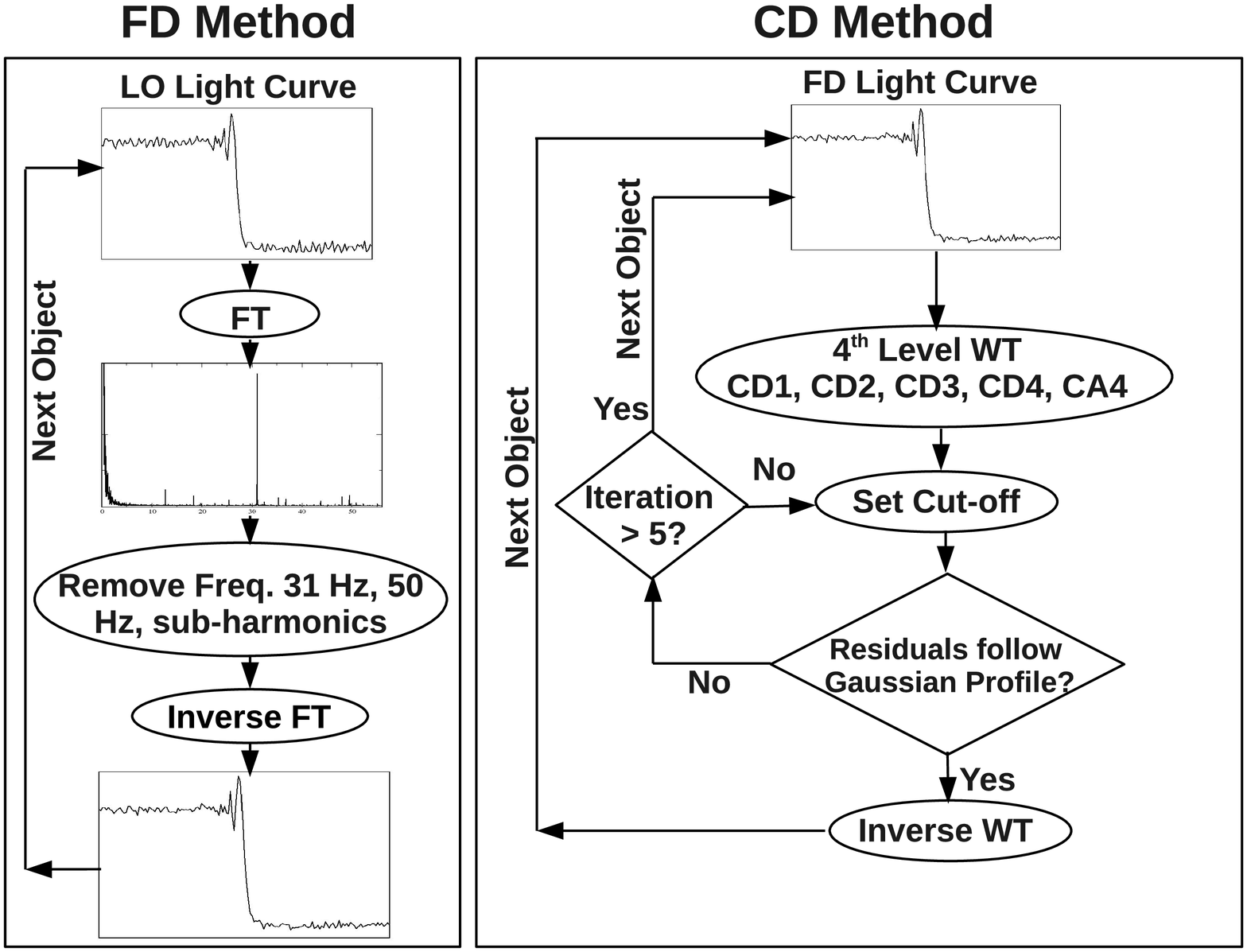}
\caption{The schematic-diagram of Fourier (left) and Combination (Fourier + Wavelet) denoising methods (right).}
\label{block_dia}
\end{figure}
The very initial and fundamental problem in applying the FT on LO data is the characteristic of FT itself. The FT of a step function (Fig.\ref{ft_step_fn}) has rapidly decreasing power ($\propto \frac{1}{freq^2}$) from low frequency to higher frequencies. While a step function with slope shows similar pattern with an additional low frequency modulation associated with it (Fig.\ref{ft_step10}). The signal level in LO light curves also changes before and after the occultation, like a step function with slope, and shows a similar characteristic in the Fourier domain (Fig.\ref{ft_of_model}). The high frequency oscillations in the power spectrum arises due to step function like nature of the light curve. The low frequency modulation in the power spectrum is obtained from the nature of falling of the light curve after occultation. Flatter the slope in light curves higher the frequency of second type of variation. This implies that even the well identified noise frequencies are not easy to remove completely, as a part of its power is always contributing to define the shape of the light curve itself. The effect is higher in the low frequency regime ($<$ 10 Hz) of the power spectrum. The frequencies greater than 10 Hz can be removed completely from the light curves without creating any significant distortion to it. A schematic diagram of the denoising method is shown in Fig.\ref{block_dia} (left).

\begin{figure}[ht]
  \centering
\includegraphics[width=14cm,height= 9cm]{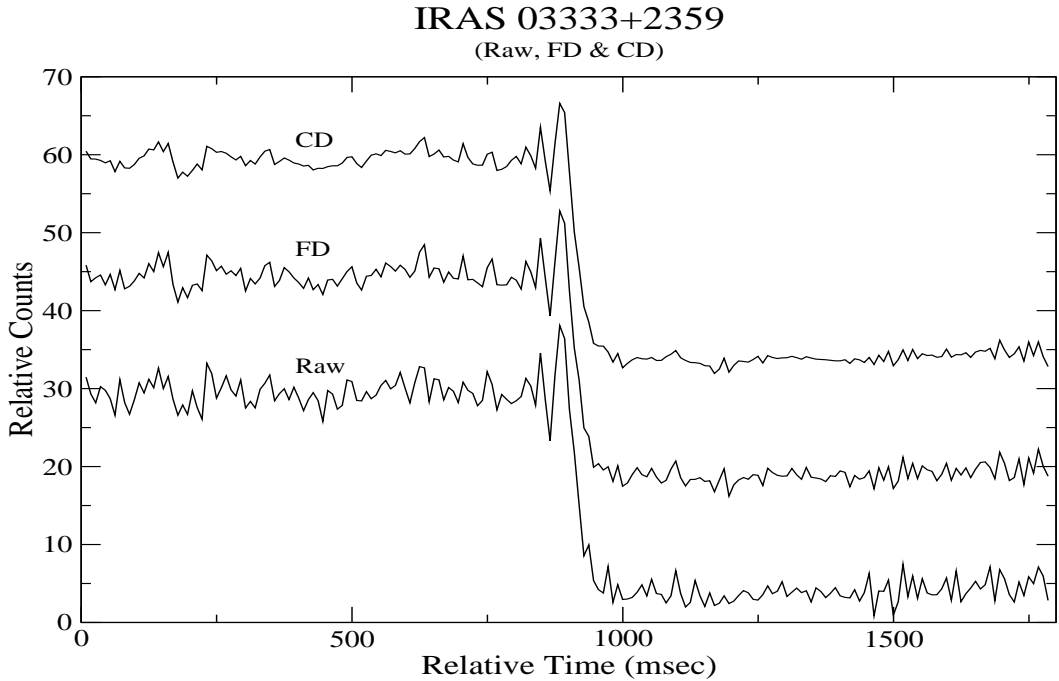}
\caption{Raw, Fourier and combination of Fourier and Wavelet denoised light curves of IRAS 03333+2359. The Fourier and Combined denoised light curves are shifted upward by 15 and 30 counts respectively to avoid overlap and for clarity. S/N has improved from 21 (`Raw') to 24 (FD) which is further improved to 32 after the application of CD method.}
\label{iras03333_time}
\end{figure}

\begin{figure*}
\includegraphics[width=14 cm,height= 8.5 cm]{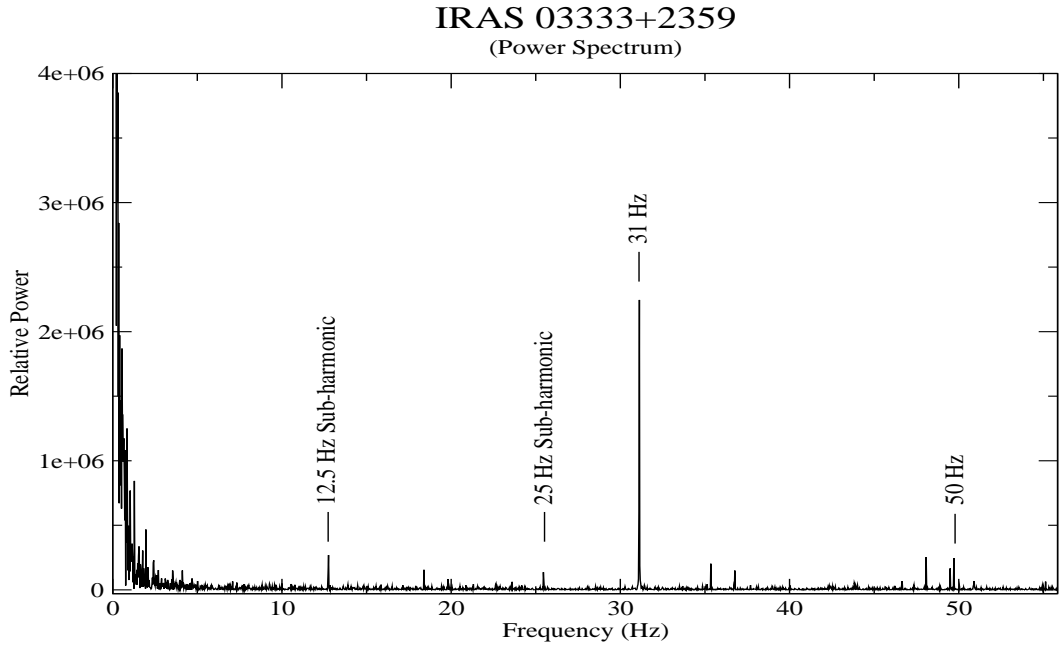}
\caption{The power spectrum of the `Raw' light curve of IRAS 03333+2359. The frequencies, marked in the spectrum, are removed from the light curve by applying the $FD$ method.}
\label{iras03333_pow_ori_ft}
\end{figure*}
\begin{figure}[ht]
\centering
\includegraphics[width=14cm,height= 8.5cm]{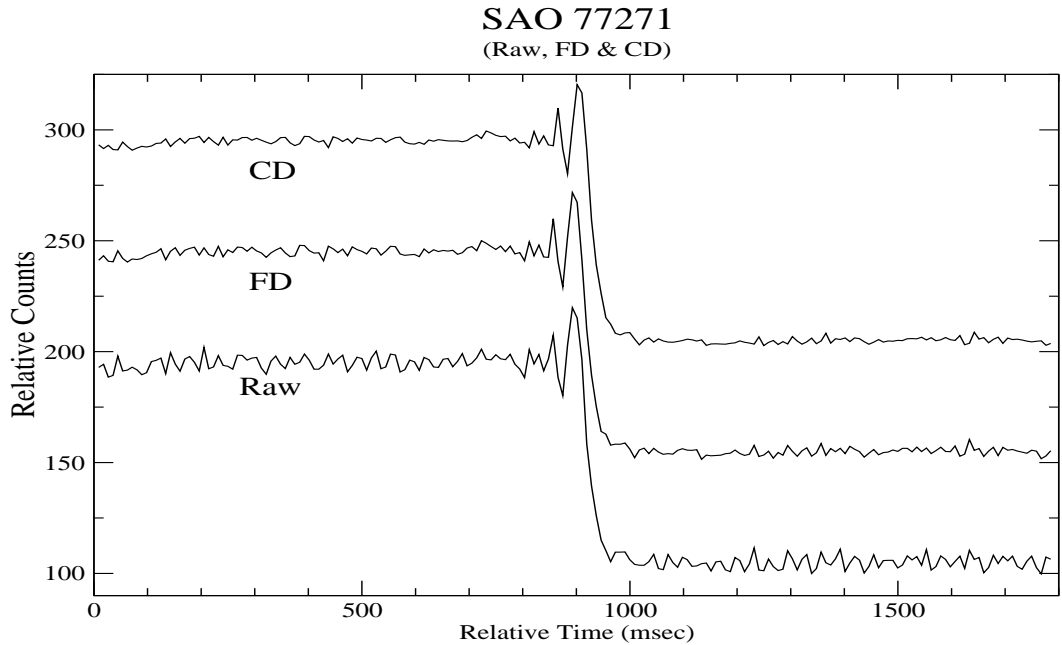}
\caption{Raw, Fourier and combination of Fourier and Wavelet denoised light curves of SAO 77271. The Fourier and combined denoised light curves are shifted upward by 50 and 100 counts to avoid overlap and for clarity. The denoised light curves can be clearly seen to have lesser noise than the observed (`Raw') data. S/N has improved from 32 (`Raw') to 55 (FD) which has improved further to 74 (CD).}
\label{sao77271_raw_fd}
\end{figure}
\begin{figure}[ht]
\centering
\includegraphics[width=14cm,height= 8.5cm]{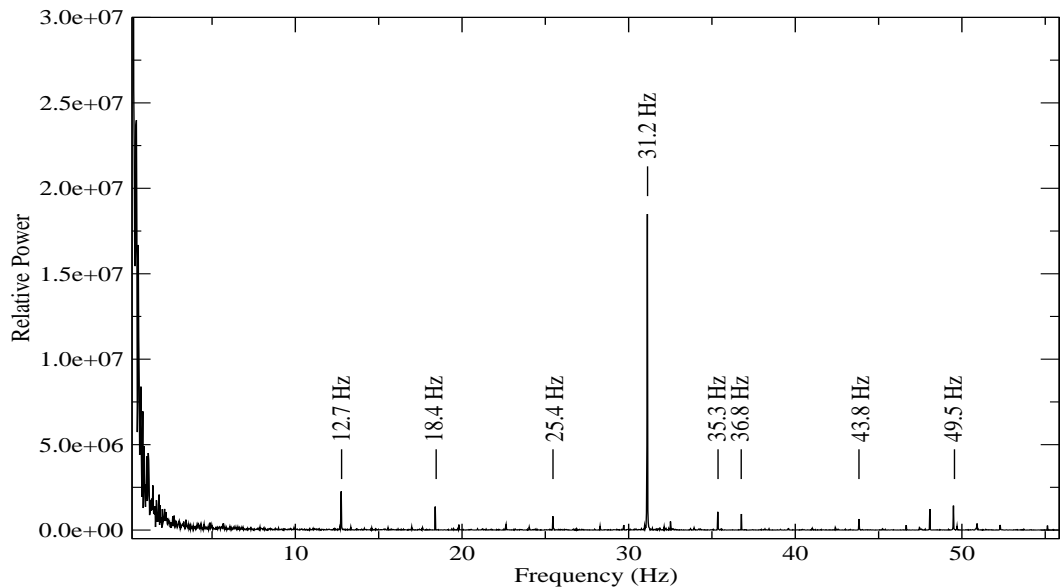}
\caption{The power spectrum of the `Raw' light curve of SAO 77271. The frequencies, marked in the spectrum, are removed from the light curve by applying the $FD$ method.}
\label{sao77271_pow}
\end{figure}

The faintest object in our sample, IRAS 03333+2359 has also shown a slight improvement after the application of FD method (Fig.\ref{iras03333_time}). The improvement in S/N has been obtained from 21 to 24. The corresponding power spectrum of the light curve is also shown in Fig.\ref{iras03333_pow_ori_ft} and removed components are marked in it. A few well identified noise frequencies, like 50 Hz power line pick-up, its sub-harmonics, and 31 Hz etc, are seen in the spectrum. It also occasionally happens that fringe frequency almost matches to the low frequency fluctuations in the signal. It has been noticed that 31 Hz noise frequency is present almost in all light curves and it contributes dominating power in many sources (mainly fainter objects). The main targets in $FD$ method were to remove these two (31 Hz \& 50 Hz) well identified noise frequencies and their sub-harmonics. But in few cases some other frequencies, with dominating power, have also been removed from the light curves but those frequencies are higher than 10 Hz. A $MATLAB$ programming script was written and used to pursue this method. The method was applied on all 54 sources listed in Table.\ref{src_list}. The light curve of the source $SAO$ $77271$ which has shown best improvement in S/N after the application of FD method (Fig.~\ref{sao77271_raw_fd}). The power-spectrum of the corresponding light curve is also shown in Fig.~\ref{sao77271_pow}. The removed frequencies are marked in the power spectrum. Removal of those frequencies enhanced the S/N of the light curve from 32 to 55.

\begin{figure*}
\includegraphics[width=14 cm,height= 8.5 cm]{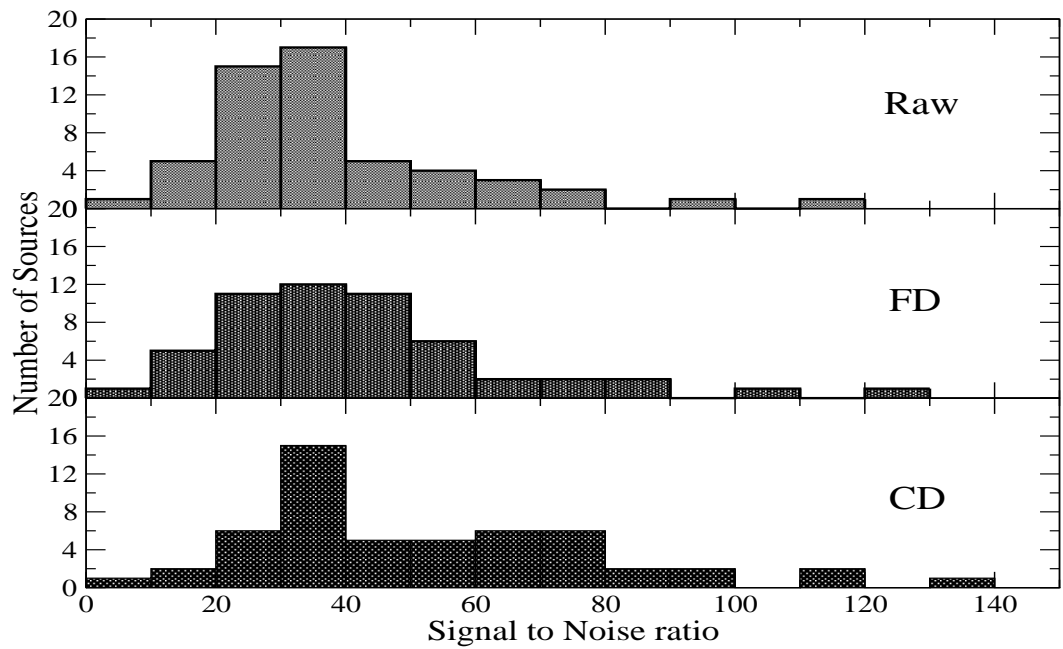}
\caption{The Histogram shows that 9 sources have moved to S/N $>$ 40 after the application of Fourier Transform method. The number has further increased by 5 after the application of Wavelet in CD method.}
\label{hist}
\end{figure*}

A histogram plot of source vs S/N for both observed ($Raw$ hereafter) (upper) and FD (lower) is displayed in Fig~\ref{hist}. It is seen that $FD$ method has improved the S/N, but not dramatically. A few light curves with lower S/N have moved towards the S/N $>$40.

With the help of Fourier transforms only few specific noise frequencies has been removed. Fluctuations for a limited period may also present in the light curves which would contain lesser power in the frequency-power domain. Hence, it is difficult to identify and remove them individually. Handling of these noisy light curves motivated us to investigate the use of wavelet transforms.

\section{WT Application on LO data}
The $MATLAB$ wavelet tools and also scripts written by us have been used to pursue the noise reduction of the LO light curves using wavelet. Many mathematical forms of wavelets are present in the literature. Generating user defined wavelet filter is also possible, subject to satisfy certain conditions. To select a suitable wavelet to be used in decomposition of LO data, a sample of 15 different wavelets on few light curves has been analyzed. It was found that two specific wavelets among them, namely, coiflet 4 and symlet 20 are suitable for filtering the LO light curves.

Application of wavelet transforms on a signal first decompose it into several details coefficients and one average coefficient subjected to the level of decomposition. For example, after 3$^{rd}$ level decomposition it produces three detail coefficients, namely, $cd1$, $cd2$, $cd3$ and one average coefficient $ca3$, where $cd1$ corresponds coefficients of highest frequencies. Photon noise contributes to the two highest frequency coefficients. The contribution from the scintillation creates low frequency variations in the light curve and hence, they come in the higher level details coefficients.

To remove the high frequency noise from the signal, cut-off limits for each components are implemented manually. Two modes of cut-off are available in MATLAB : (1) Soft-cutoff \& (2) Hard-cutoff. In `Soft-cutoff' a certain amount of amplitude is deducted from all the coefficient, while in the case of `Hard-cutoff' amplitude above the given cutoff value are clipped. Here, `Soft-cutoff' has been used for noise removal of the light curves. In LO light curves mainly the highest frequency coefficients are removed and they are Gaussian noise. Hence, the condition imposed that their residuals (Raw $-$ denoised) should follow a Gaussian distribution with zero mean.  A schematic diagram of the denoising method is shown in Fig.\ref{block_dia} (right).

Wavelet denoising has been applied on all 54 LO light curves. Though significant improvement was obtained in the S/N using Wavelet transform method, it is found that purely noise frequencies like 31 Hz, 50 Hz are also present with significant power in the wavelet denoised light curves. The wavelet denoising method also has an averaging effect, because it deducts the detail coefficients with a certain cutoff. The reduction of high fluctuation from the average base-level would always lead to a smoothening effect. Application of wavelet transforms on lunar occultation light curve acts as running average and finally smoothen the observed light curves. In the model fit, averaged fringes fit to higher uniform disk (UD) angular diameter values than the source originally is. 
\begin{figure*}
\includegraphics[width=14 cm,height= 8.5 cm]{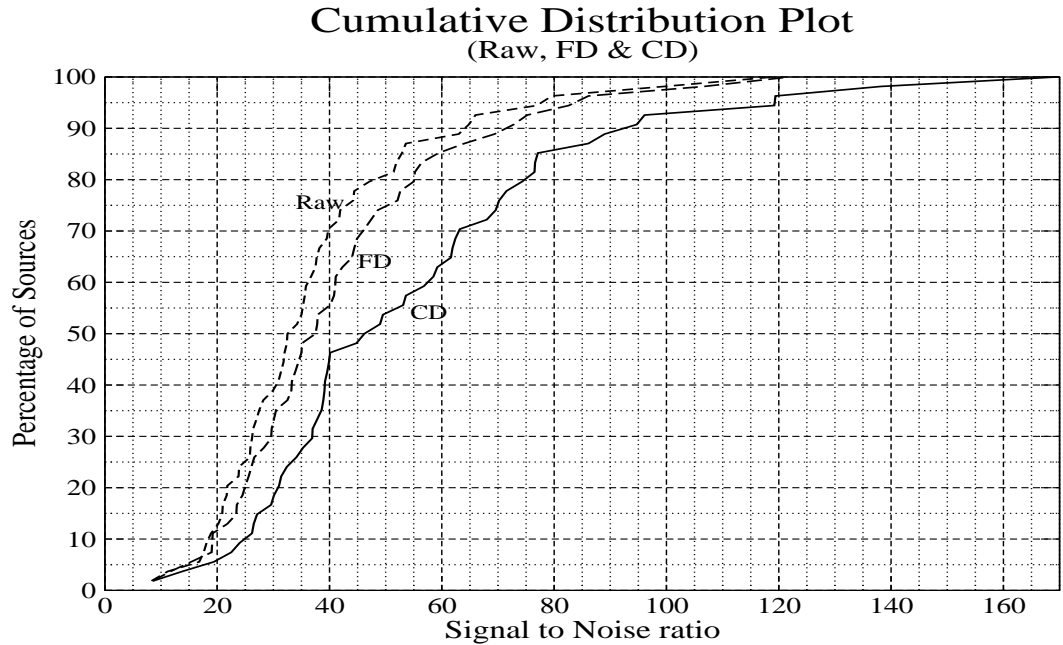}
\caption{The dotted line is the cumulative distribution plot of Raw light curves. It has been shifted towards right hand-side (solid line) after the application of denoising methods. For example before application of the denoising method, 70\% sources had S/N $<$ 40. After the application of the method only 56\% of the sources have S/N $<$ 40 and it has further decreased to 45\% after application of WT in CD method.}
\label{cumulative}
\end{figure*}

So, it was decided to use a combination of both the methods. First, $FD$ is carried out to remove those well identified specific noise frequencies. Then WT method is applied in similar manner as explained above. The WT method is executed after the application of $FD$ to put minimum cut-offs on WT coefficients. This is because the averaging effect is cut-off dependent. Hence, the combination of Fourier and Wavelet denoising ($CD$) method has been applied on all 54 LO light curves by setting the cut-off limit manually. Here also we imposed the same condition that residuals (Raw $-$ Denoised) should follow Gaussian distribution with zero mean.

\section{Results \& Discussion}
The result of the applied methods has two aspects : (1) the improvement in the S/N of the light curves and (2) improvement in the model fits for possible angular diameter determination. We first discuss enhancements in S/N using the two methods, namely, $FD$ and $CD$ methods. The numerical details before and after the application of denoising methods are listed in Table.\ref{result_table}. The table here is arranged in the increasing order of RA. The first column of the table lists the source designations and the second column is the apparent K-magnitude as given in 2MASS catalog. The S/N of the Raw and denoised light curves using $FD$ and $CD$ methods are listed in next three columns of the table respectively.
\subsection{Improvement in S/N after denoising methods}
\subsubsection{S/N after $FD$ method}
The improvement in the S/N using the $FD$ method is graphically shown in the histogram plot in Fig.\ref{hist}. It shows that 9 out of 38 sources have improved their S/N beyond 40 after application of $FD$ method which implies that $~$24\% improvement has been obtained. A different cumulative representation of the same result is shown in Fig.\ref{cumulative}. Here, the X-axis corresponds to the S/N of the light curves and Y-axis represents the percentage of sources below any specific S/N. For example the dotted line, representing `Raw' light curves, shows that 70\% of the sources have S/N below 40. It reduces to 56\% after the application of $FD$ method (slashed line).

The sources which show significant S/N improvement are marked with bold face in column 4 of Table.\ref{result_table}. It can be noticed that a total of 29 out of 54 sources (more than 50\%) show significant ($\geq$ 3$\sigma$) improvement in S/N using the $FD$ method. The remaining 25 sources did not show any improvement in their S/N. Out of these 25 sources 14 are brighter than $m_K \sim$ 3.0. Noise in these light curves is dominated by low frequency ($<$ 5 Hz) noise. But the main noise components removed in the $FD$ method are 50 Hz, its sub-harmonics (25 Hz \& 12.5 Hz) and 31 Hz. The removal of low frequency noise (mainly due to scintillation) is difficult as it is not at a single frequency but arise upto 5 Hz (see Fig.\ref{sao118044_ft}) with varying power. Complete removal of low frequency component leads to distortion of LO fringes. Hence, S/N of the 14 brighter sources could not be improved by $FD$ method.

Out of the remaining 25 sources 11 are fainter than $m_K \sim$ 3.0. It is found that 9 of those 11 sources were observed in the presence of high speed wind ($>$ 40 km/hr) resulting in telescope oscillations (Frequency $<$ 5 Hz). One source (SAO 145698) was observed during day time when back-ground noise was very high. Another source (SAO 76450) occasionally drifted out of the 10$\times$10 pixels sub-array due to improper telescope tracking at lower elevation. Hence, these 11 fainter sources did not show S/N improvement due to observational difficulties.

The average enhancement obtained in the value of S/N for 29 sources is 18\% after application of $FD$ method. It is also observed that among the 29 sources which show significant S/N improvement, 19 are fainter than $m_K \sim$3. Infact the largest improvement in S/N using $FD$ method is obtained in the light curve of a faint source SAO 77271 ($m_K \sim$3.3). S/N of the light curve has improved from 32 to 55 (72\%). The fainter objects ($m_K > $ 3.0) are mainly affected by the sky background photon noise but have a higher noise content at specific frequencies (31 Hz, 50 Hz etc) which is removed by $FD$ method. That is why, they show a good response to the application of the $FD$ method on removing noise at specific frequencies (31 Hz, 50 Hz etc).
\subsubsection{S/N after $CD$ method}
We now take up S/N improvement obtained using the combination method. It must be noted that in $CD$ method, $FD$ is applied first, followed by the wavelet transform. Wavelet transforms decompose any signal into `average coefficients' and `detail coefficients'. It is possible to directly remove fluctuations from the light curve without knowledge of their frequencies by setting suitable cutoffs. That is why it is not surprising that $CD$ method has shown improvement in S/N of all but five of the light curves significantly (except IRC -10564, RT Cap, IRAS 04320+2519$^a$, SAO 92755$^b$ \& CW Cnc) and performed better than $FD$ method. The sources IRC -10564 \& CW Cnc were observed at lower elevation of 16$^o$ and 18$^o$ respectively. Their light curves are affected by large scintillations. RT cap was observed in broad day light and its light curve is affected by large back-ground noise. It was partly cloudy during the observation of IRAS 04320+2519$^a$ and telescope oscillation added uneven noise in the light curve of SAO 92755$^b$. Hence, denoising methods did not yield improvement for these cases.

It is mentioned above that in the CD method, wavelet transform is applied after the application of $FD$ method on observed light curves. So, the S/N of 29 light curves is already enhanced by average of 18\% using $FD$ method before application of wavelet. The histogram plot (Fig.\ref{hist}) shows that 38 sources were below S/N $\sim$40 before application of any denoising method. The number reduced to 29 by $FD$ method which has further reduced to 24 after the application of wavelet method. The cumulative histogram plot in Fig.\ref{cumulative} (dotted curve) shows that 70\% of the `Raw' light curves fall in the region S/N $<$ 40. The application of $FD$ method has reduced it to 56\% (slashed curve). Then the application of wavelet transforms (in $CD$) has further reduced it to 45\% (solid curve). The faintest object in our sample, IRAS 03333+2359, which has shown S/N improvement from 21 to 24 after $FD$ method, has further shown improvement to S/N of 32 after the application of CD method (Fig.\ref{iras03333_time}). One source (SAO 77271) has shown a significant improvement after application of $FD$ and $CD$ methods. It has shown improvement in S/N from 32 to 55 after $FD$ method which is further improved to 74 after the application of wavelet transform in $CD$ method (Fig.\ref{sao77271_raw_fd}).
\subsection{Model fits to denoised light curves}
We next consider the possible improvement in the light curves for better model fits to determine the uniform disk (UD) values (or UD limit on unresolved sources) after the application of $FD$ and $CD$ methods. To examine this aspect, fits to all three (i.e., $Raw$, $FD$ \& $CD$) light curves of all the sources have been carried out. The light curves were fitted using non-linear least square method. Details on the model fit parameters and analysis can be found in Chandrasekhar \& Baug (2010).

Out of the original sample of 54 sources considered for S/N improvement, 50 sources could be used for model fits. The remaining four sources though bright ($m_K <$3.0), could not be fitted to a model curve due to the following reasons. The light curve of SAO 92755$^b$ is too noisy to fit due to telescope tracking problems and noise could not be reduced by FD or CD methods. The light curves of TU Gem and VV Cnc suffer from fringe distortions possibly because of variable rate of lunar component velocity due to limb irregularities. The light curve of IRC +20177 was observed in 20$\times$20 pixels sub-array to accommodate its visual binary component. Sampling time in 20$\times$20 pixels sub-array is poorer than 10$\times$10. Its poor sampling time of $\sim$14 msec, provided a few data points on the fringes and could not be fitted properly. However its binary separation of 700$\pm$50 mas is obtained at a position angle of 109$^o$.

After obtaining the best fit to the light curves, the Standard Deviation (SD) of the residuals (data $-$ model) have been calculated for ($a$) total data set of 200 points ($\sim$1.8 sec) used in the model fit and ($b$) fringe portion (40 data points with first fringe of the light curve in the middle). Ideally values of standard deviation in the fringe portion (SD$_{fringe}$) as well as total data set (SD$_{total}$) should decrease after the application of denoising methods. We expect this to happen because noise in the observed light curve are reduced by the denoising methods. It was indeed found that the SD$_{fringe}$ and SD$_{total}$ values for denoising light curves are lower than the `Raw' data in most of the sources. However, we have introduced one more parameter called SD$_{ratio}$ (=$^{SD_{fringe}}$/$_{SD_{total}}$) to examine the improvement of the model fit in the fringe region relative to the rest of the light curve. SD$_{ratio}$ along with SD$_{fringe}$ and SD$_{total}$ values for all three cases 
($Raw$, $FD$ \& $CD$) are listed in columns 6-14 of Table.\ref{result_table}. Lower the value of SD$_{ratio}$ better is the fit to the fringes.
\begin{center}
\begin{longtable}{l|c|ccc|ccc|ccc|ccc}
\caption{Numerical details of $Raw$, $FD$ and $CD$ light curves.}
\label{result_table}\\
\hline
Source & $m_K$ & \multicolumn{3}{|c|}{S/N} & \multicolumn{3}{|c|}{SD$_{fringe}$} & \multicolumn{3}{|c|}{SD$_{total}$} & \multicolumn{3}{|c}{SD$_{ratio}$} \\
       &       &  Raw   &   FD   &   CD    &    Raw     &     FD    &     CD     &    Raw    &     FD     &     CD    &     Raw    &	   FD    &    CD   \\
\endfirsthead

\multicolumn{14}{c}{{\tablename\ \thetable{} -- continued from previous page}}\\
\hline
Source & $m_K$ & \multicolumn{3}{|c|}{S/N} & \multicolumn{3}{|c|}{SD$_{fringe}$} & \multicolumn{3}{|c|}{SD$_{total}$} & \multicolumn{3}{|c}{SD$_{ratio}$} \\
       &       &  Raw   &   FD   &   CD    &    Raw     &     FD    &     CD     &    Raw    &     FD     &     CD    &     Raw    &	   FD    &    CD   \\
\hline
\endhead

\hline
\endfoot

\hline
SAO 109252		& 		3.74 		&		31 		&		\textbf{35} 		&		\textbf{53} 		&		0.64 		&		0.52 		& 		0.45		&		0.66		&		0.59		&		0.42		&		0.97		& 		0.87		&		1.06		\\
SAO 92697		& 		2.03 		&		47 		&		\textbf{52}		&		\textbf{62}		&		0.92		&		0.81		&		0.71		&		1.02		&		0.89		&		0.71		&		0.90		&		0.91		&		0.99		\\
SAO 92755$^a$	      	&		2.50		&		28		&			30	      	&		\textbf{31}		&		1.25		&		1.51		&		1.43		&		1.80		&		1.78		&		1.68		&		0.70		&		0.85		&		0.85		\\
SAO 92755$^b$	      	&		2.50		&		09		&			09	      	&			09		&		--		&		--		&		--		&		--		&		--		&		--		&		--		&		--		&		--		\\
IRC +20037	    	&		2.90		&		34		&			35	      	&		\textbf{40}		&		1.52		&		1.51		&		1.38		&		1.89		&		1.84		&		1.61		&		0.80		&		0.82		&		0.86		\\
UY Ari       		&		3.69		&		37		&		\textbf{41}		&		\textbf{70}		&		0.90		&		0.84		&		0.68		&		0.74		&		0.67		&		0.45		&		1.22		&		1.26		&		1.53		\\
AU Ari       		&		3.68		&		24		&			25	      	&		\textbf{34}		&		0.97		&		0.89		&		0.83		&		0.89		&		0.83		&		0.60		&		1.09		&		1.08		&		1.38		\\
SAO 75669	      	&		3.35		&		26		&		\textbf{30}		&		\textbf{45}		&		0.83		&		0.65		&		0.56		&		0.94		&		0.83		&		0.52		&		0.88		&		0.79		&		1.08		\\
UZ Ari$^a$       	&		1.25		&		42		&			42	      	&		\textbf{46}		&		6.67		&		7.08		&		6.83		&		4.88		&		4.88		&		4.52		&		1.37		&		1.45		&		1.51		\\
UZ Ari$^b$	    	&		1.25		&		77		&		\textbf{83}		&		\textbf{95}		&		2.22		&		2.09		&		1.92		&		1.80		&		1.73		&		1.56		&		1.24		&		1.21		&		1.24		\\
SAO 75706	      	&		3.71		&		18		&			19	      	&		\textbf{23}		&		0.84		&		0.86		&		0.82		&		0.92		&		0.84		&		0.67		&		0.91		&		1.03		&		1.22		\\
IRAS 03333+2359		&		5.07		&		21		&		\textbf{24}		&		\textbf{32}		&		0.67		&		0.53		&		0.41		&		0.71		&		0.58		&		0.41		&		0.94		&		0.90		&		0.98		\\
IRC +20066	    	&		2.60		&		65		&		\textbf{75}		&		\textbf{119}		&		1.05		&		1.01		&		1.05		&		0.98		&		0.88		&		0.68		&		1.08		&		1.15		&		1.53		\\
NSV 1406	      	&		2.93		&		26		&			26	      	&		\textbf{30}		&		1.22		&		1.08		&		1.00		&		1.16		&		1.08		&		0.89		&		1.05		&		1.00		&		1.12		\\
SAO 76350	      	&		3.20		&		66		&		\textbf{69}		&		\textbf{96}		&		0.90		&		0.86		&		0.87		&		0.91		&		0.84		&		0.68		&		0.99		&		1.02		&		1.28		\\
V1134 Tau   		&		3.84		&		40		&			40	      	&		\textbf{54}		&		1.03		&		1.00		&		0.88		&		0.95		&		0.91		&		0.70		&		1.09		&		1.10		&		1.26		\\
SAO 76450	      	&		3.06		&		21		&			22	      	&		\textbf{27}		&		1.16		&		1.16		&		1.08		&		1.06		&		1.02		&		0.83		&		1.10		&		1.14		&		1.30		\\
IRAS 04320+2519		&		4.30		&		18		&			19	      	&		\textbf{24}		&		0.94		&		0.88		&		0.75		&		0.89		&		0.81		&		0.66		&		1.05		&		1.09		&		1.12		\\
IRAS 04320+2519		&		4.30		&		11		&			12	      	&		\textbf{14}		&		1.44		&		1.37		&		1.21		&		1.15		&		1.08		&		0.89		&		1.25		&		1.27		&		1.36		\\
C* 3246       		&		3.58		&		44		&		\textbf{59}		&		\textbf{86}		&		0.87		&		0.67		&		0.55		&		0.78		&		0.65		&		0.47		&		1.11		&		1.03		&		1.17		\\
IRAS 05013+2704		&		4.12		&		35		&		\textbf{41}		&		\textbf{62}		&		0.71		&		0.64		&		0.55		&		0.69		&		0.61		&		0.42		&		1.03		&		1.05		&		1.32		\\
SAO 76952	      	&		4.70		&		17		&			19	      	&		\textbf{27}		&		0.81		&		0.71		&		0.60		&		0.93		&		0.82		&		0.61		&		0.87		&		0.86		&		0.98		\\
SAO 76965	      	&		4.40		&		28		&		\textbf{31}		&		\textbf{39}		&		0.66		&		0.57		&		0.49		&		0.67		&		0.58		&		0.45		&		0.99		&		0.98		&		1.08		\\
IRAS 05212+2655		&		3.74		&		26		&			28	      	&		\textbf{31}		&		1.81		&		1.68		&		1.60		&		1.48		&		1.38		&		1.25		&		1.22		&		1.22		&		1.28		\\
IRAS 05232+2400		&		3.94		&		33		&			34	      	&		\textbf{40}		&		1.91		&		1.94		&		1.76		&		1.64		&		1.60		&		1.46		&		1.17		&		1.21		&		1.20		\\
SAO 77229	      	&		3.87		&		32		&		\textbf{38}		&		\textbf{49}		&		0.96		&		0.90		&		0.87		&		1.00		&		0.86		&		0.67		&		0.96		&		1.05		&		1.30		\\
SAO 77271	      	&		3.27		&		32		&		\textbf{55}		&		\textbf{74}		&		1.35		&		0.90		&		0.86		&		1.42		&		0.85		&		0.68		&		0.95		&		1.06		&		1.25		\\
SAO 77357	      	&		4.39		&		22		&			27	      	&		\textbf{40}		&		0.67		&		0.55		&		0.48		&		0.60		&		0.52		&		0.36		&		1.11		&		1.06		&		1.34		\\
SAO 77474       	&		4.14		&		22		&		\textbf{25}		&		\textbf{35}		&		0.64		&		0.62		&		0.54		&		0.73		&		0.67		&		0.50		&		0.88		&		0.94		&		1.08		\\
HD 249571	      	&		4.50		&		26		&		\textbf{31}		&		\textbf{37}		&		0.73		&		0.58		&		0.51		&		0.84		&		0.74		&		0.59		&		0.88		&		0.78		&		0.87		\\
SAO 77792	      	&		2.97		&		40		&		\textbf{45}		&		\textbf{62}		&		1.36		&		1.22		&		1.11		&		1.03		&		0.92		&		0.73		&		1.32		&		1.33		&		1.52		\\
IRAS 05551+2305		&		4.18		&		44		&		\textbf{47}		&		\textbf{70}		&		0.96		&		0.85		&		0.84		&		0.85		&		0.82		&		0.61		&		1.13		&		1.03		&		1.36		\\
BI Gem       		&		3.17		&		52		&		\textbf{64}		&		\textbf{89}		&		0.74		&		0.66		&		0.62		&		0.81		&		0.65		&		0.50		&		0.91		&		1.01		&		1.24		\\
IRC +30140	    	&		2.70		&		100		&		\textbf{107}		&		\textbf{169}		&		1.07		&		0.91		&		0.94		&		0.87		&		0.77		&		0.53		&		1.22		&		1.19		&		1.78		\\
TU Gem		    	&		0.82		&		32 		&		33          		&		\textbf{38}		&		--		&		--		&		--		&		--		&		--		&		--		&		--		&		--		&		--		\\
BD+26 1131	    	&		4.06		&		38		&		\textbf{46}		&		\textbf{68}		&		0.78		&		0.57		&		0.50		&		0.77		&		0.66		&		0.45		&		1.02		&		0.87		&		1.11		\\
IRAS 06165+2431		&		4.87		&		20		&		\textbf{23}		&		\textbf{30}		&		0.63		&		0.55		&		0.47		&		0.64		&		0.56		&		0.45		&		0.97		&		0.98		&		1.06		\\
IRAS 06395+2409		&		3.66		&		42		&		\textbf{49}		&		\textbf{77}		&		0.86		&		0.73		&		0.60		&		0.78		&		0.68		&		0.46		&		1.10		&		1.07		&		1.29		\\
IRC +20177   	      	&		2.56		&		30		&		\textbf{41}		&		\textbf{50}		&	--			&		--		&		--		&		--		&		--		&		--		&		--		&		--		&		--		\\
IRC +20186	    	&		2.18		&		35		&			35	      	&		\textbf{39}		&		4.76		&		4.69		&		5.06		&		3.68		&		3.71		&		3.61		&		1.30		&		1.26		&		1.40		\\
VV Cnc       	      	&		0.94		&		32		&			32	      	&		\textbf{37}		&	--			&		--		&		--		&		--		&		--		&		--		&		--		&		--		&		--		\\
CW Cnc       		&		0.11		&		19		&			19	      	&			19	      	&		7.43		&		8.08		&		7.92		&		5.49		&		5.86		&		5.67		&		1.35		&		1.38		&		1.40		\\
6 Leo       		&		1.98		&		54		&			55	      	&		\textbf{63}		&		5.31		&		5.34		&		5.24		&		4.16		&		4.10		&		3.83		&		1.29		&		1.30		&		1.37		\\
IRC +10210	    	&		2.50		&		38		&		\textbf{44}		&		\textbf{57}		&		1.12		&		1.08		&		1.00		&		1.03		&		0.91		&		0.72		&		1.09		&		1.18		&		1.39		\\
SAO 118044	    	&		0.49		&		119		&		\textbf{122}		&		\textbf{138}		&		2.48		&		2.44		&		2.37		&		2.07		&		2.03		&		1.87		&		1.20		&		1.20		&		1.27		\\
IRC +00202	    	&		2.59		&		53		&		\textbf{56}		&		\textbf{77}		&		0.87		&		0.82		&		0.64		&		0.78		&		0.70		&		0.54		&		1.12		&		1.17		&		1.19		\\
SAO 157613	    	&		3.24		&		36		&		\textbf{44}		&		\textbf{72}		&		0.67		&		0.65		&		0.67		&		0.75		&		0.66		&		0.52		&		0.89		&		0.99		&		1.29		\\
CW Oph       		&		2.65		&		27		&		\textbf{33}		&		\textbf{39}		&		1.44		&		1.16		&		0.98		&		1.54		&		1.29		&		1.10		&		0.93		&		0.90		&		0.90		\\
RT Cap       		&		0.31		&		38		&			38	      	&			39	      	&		11.16		&		11.27		&		10.87		&		9.76		&		9.79		&		9.47		&		1.14		&		1.15		&		1.15		\\
RS Cap       		&		-0.22		&		51		&			53	      	&		\textbf{59}		&		4.25		&		3.96		&		3.86		&		3.55		&		3.48		&		3.33		&		1.20		&		1.14		&		1.16		\\
IRC -10564	    	&		2.22		&		24		&			26	      	&			  26	    	&		2.78		&		2.77		&		2.72		&		2.81		&		2.79		&		2.76		&		0.99		&		0.99		&		0.98		\\
SAO 145698	    	&		3.80		&		36		&			38	      	&		\textbf{59}		&		0.58		&		0.54		&		0.44		&		0.59		&		0.54		&		0.36		&		0.99		&		1.00		&		1.20		\\
SAO 145992	    	&		3.22		&		80		&		\textbf{86}		&		\textbf{119}		&		0.87		&		0.85		&		0.77		&		0.96		&		0.91		&		0.75		&		0.91		&		0.93		&		1.04		\\
TX Psc       		&		-0.51		&		63		&		\textbf{73}		&		\textbf{77}		&		1.53		&		1.19		&		0.90		&		1.29		&		1.03		&		0.83		&		1.18		&		1.15		&		1.09		\\
\end{longtable}
\end{center}
\subsection{Improvement in model fits}
\begin{figure*}
\includegraphics[width=14 cm,height= 10 cm]{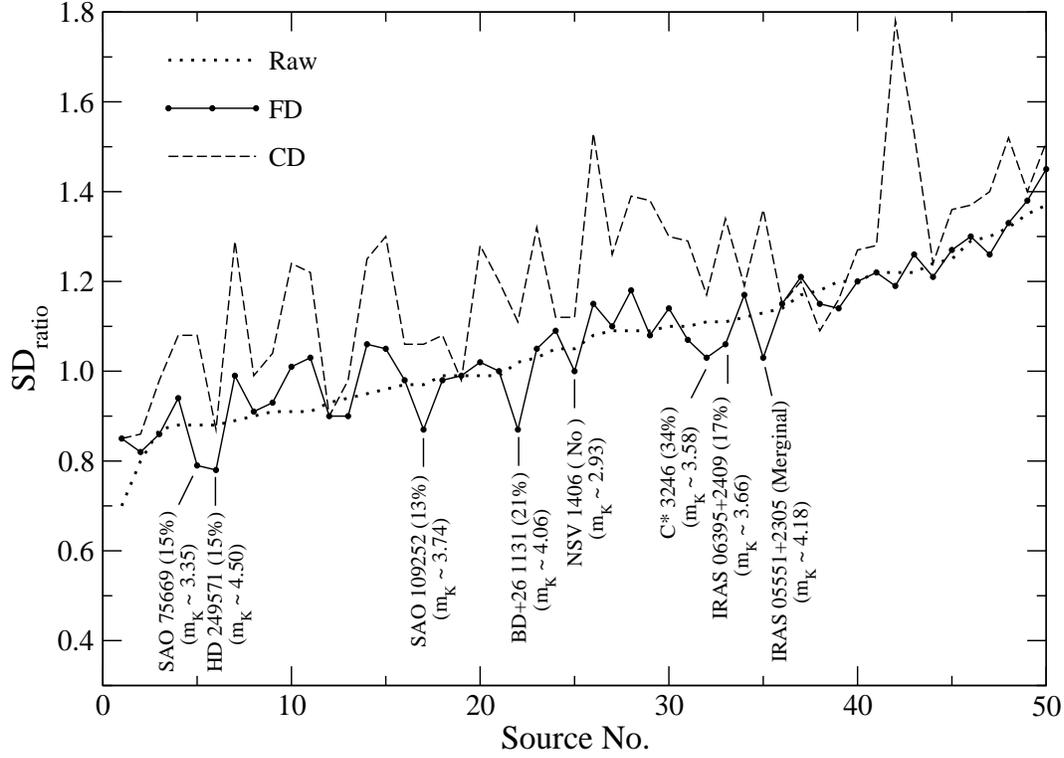}
\caption{SD$_{ratio}$ of $Raw$, $FD$ and $CD$ light curves are plotted with SD$_{ratio}$ of $Raw$ data arranged in ascending order. The SD$_{ratio}$ values of $CD$ methods are generally higher (poorer fit to the light curve) than the other two. $SD_{ratio}$ for $FD$ fluctuates with respect to the $Raw$ values. The sources which show lowest SD$_{ratio}$ values for $FD$ method (best fit to the light curves) are marked along with their S/N enhancement (in percentage) obtained using $FD$ method.}
\label{sd_ratio}
\end{figure*}

When we examine Table.\ref{result_table}, in most of the cases SD$_{ratio}$ for $CD$ is higher than SD$_{ratio}$ for `Raw' and $FD$ light curves. It shows that $CD$ light curves fit poorly to fringes compared to $FD$ or $Raw$ light curves. This result arises because the S/N enhancement seen in $CD$ method is mainly due to better denoising effect in the non-fringe region compared to the fringe region. $CD$ method results in a slight smoothening of the light curves in general and leads to a higher UD value in the model fit. Only in the case of a well resolved source (TX Psc), SD$_{ratio}$ value for $CD$ is lower than other two. These results are also illustrated in Fig.\ref{sd_ratio}. In the figure, SD$_{ratio}$ values of all three methods are plotted arranging the SD$_{ratio}$ values for $Raw$ data in ascending order. It can be seen that while the SD$_{ratio}$ values for $FD$ method is fluctuates above and below with respect to the $Raw$ values, the ratio generally show higher values for $CD$ than other two methods.

Out of the 50 sources selected for the model fits about one third (17) have lower SD$_{ratio}$ value for the `Raw' light curve than that for $FD$ or $CD$. For half of the sources (24) the SD$_{ratio}$ is not significantly different between `Raw' and $FD$ but they are lower than $CD$ (Fig.\ref{sd_ratio}). For the remaining 8 sources; namely, BD+26 1131, C* 3246, IRAS 06395+2409, SAO 109252, SAO 75669, HD 249571, NSV 1406 \& IRAS 05551+2305, SD$_{ratio}$ value for $FD$ is significantly lower than both $Raw$ and $CD$ data. Out of the 8 sources, 6 have shown earlier significant S/N improvement on applying the $FD$ method. Those sources are marked in Fig.\ref{sd_ratio} along with their S/N improvement in percentage. These sources are all fainter than $m_K \sim$3.3.
\subsection{Resolved and unresolved sources in the sample} 
Out of 50 sources, we find on analysis that, five are reported resolved sources. Our values agree well with the earlier reported UD angular diameters. They are listed in Table.\ref{ud_resolved} along with the references to their earlier reports in the last column. A close examination of our model fits and details of sources in the literature reveals that two more sources are expected to be resolved. There are two LO observations of $UZ$ $Ari$ (IRC +20052). This source is a semi-regular variable with a periodicity of 163 days. Fits to the two light curves of the source in two different events provide comparable UD values of 5.5$\pm$0.5 mas and 6.0$\pm$0.5 mas respectively. According to the angular diameter prediction method discussed in Baug \& Chandrasekhar (2012) expected UD value for this source is 4.2$\pm$1.0 which is a little lower than the measured values. But presence of a circumstellar shell may lead to a higher UD angular diameter at 2.2 $\mu m$. The other source $SAO$ $92697$ is also a semi-regular 
variable with period of 206 days. A close examination of the model fit reveals that the source is expected to be resolved. These sources are also listed in Table.\ref{ud_resolved}.

In three resolved sources (CW Cnc, RT Cap \& UZ Ari$^a$) there is no improvement in S/N by FD method and hence, `Raw' data is the best in model fits. In two cases (RS Cap \& UZ Ari$^b$) SD$_{ratio}$ values for FD is the lowest and hence, fit best to the model. In case of RS Cap, though no S/N improvement has been obtained but the removal of noise frequencies using FD has cleaned the fringes to fit better. In the case of UZ Ari$^b$ S/N improvement using the FD method has been obtained. In two other sources (SAO 118044 \& SAO 92697) `Raw' and $FD$ data give comparable fits though S/N improvement is not seen after the application of FD method. For the important well resolved source, TX Psc, best fit is obtained by the CD method.

The light curve and model fits of the remaining sources have been carefully examined to see if any of these sources is resolved. We find all these sources to be unresolved. But the limit of resolution varies from source to source depending on the S/N and other conditions. The faintest source $IRAS$ $03333+2359$ ($m_K$ = 5.1) has a S/N$\sim$24, is best fitted with $FD$ light curve giving a limit $\leq 4.3$ mas. The lowest limit to the resolution of $\leq 3.5$ mas is obtained for UY Ari ($m_K$ = 3.7). A good S/N improvement is obtained by applying the denoising methods on this source and all three light curves fits to UD limits. For most of the sources with good light curves, we obtained a resolution limit close to $\leq 4.0$ mas.
\begin{center}
\begin{table}
\caption{Resolved sources in the sample (UD values).}
\begin{tabular}{lcccl}
\hline
    Source &   Filter  &UD Ang.    & Earlier Reported  & Reference  \\
           & ($\mu m$) &Dia. (mas) &    UD (mas)&            \\
\hline
   RT Cap  & 2.20/0.40 &8.1$\pm$0.3& 7.72$\pm$0.16 & White \& Feierman, 1987           \\
           &           &           & 8.18$\pm$0.21 & van Belle, Thompson \& PTI, 2000  \\
           &           &           &               &                                   \\
   RS Cap  & 2.37/0.10 &7.7$\pm$0.5& 7.75$\pm$0.67 & Richichi et al., 1992             \\
           &           &           & 7.70$\pm$0.80 & Dyke, van Belle \& Thompson, 1998 \\
           &           &           &               &                                   \\
   CW Cnc  & 2.20/0.40 &7.0$\pm$0.5& 7.05$\pm$0.33 & White \& Feierman, 1987           \\
           &           &           &               &                                   \\
SAO 118044 & 2.20/0.40 &4.9$\pm$0.5& 4.85$\pm$0.23 & White \& Feierman, 1987           \\
           &           &           &               &                                   \\
   TX Psc  & 2.37/0.10&10.6$\pm$0.5& 7.50 to 11.45 & Richichi et al., 2005             \\
           &           &           &               &              \\
UZ Ari$^a$ & 2.20/0.40 &6.0$\pm$0.5&      --       &    --        \\
           &           &           &               &              \\
UZ Ari$^b$ & 2.20/0.40 &5.5$\pm$0.5&      --       &    --        \\
           &           &           &               &              \\
SAO 92697  & 2.20/0.40 &4.7$\pm$0.5&      --       &    --        \\
\hline
\end{tabular}
\label{ud_resolved}
\end{table}
\end{center}
\section{Conclusions}
Noise reduction method on Lunar Occultation light curves using Fourier, Wavelet transforms and their combination has been carried out. A total of 54 light curves have been chosen as sample. We reject the Wavelet Transform alone as it smoothen the light curve which finally leads to fit improper angular diameter value. S/N enhancement occurs in both Fourier and combination methods. Combination method shows good performance in terms of S/N improvement. In a few cases Fourier denoised ($FD$) light curves which have significant S/N improvement, show best model fits. For most of the bright sources the $Raw$ data appears to be the best for model fits. It has been noted that $FD$ method performs better in the fainter regime ($m_K\geq$3.3). For fainter objects ($m_K > $ 3.3) where back-ground noise dominates, it appears that the better fit is obtained with $FD$ light curves because noise at specific frequencies are filtered out. In application of combined denoised method ($CD$) though there is substantial S/N improvement in most of the cases, but an averaging effect affects the light curves. This leads to a larger limit to the fitted uniform disk angular diameter values compared to $FD$ and $Raw$ data. However, in case of well resolved sources like TX Psc where fit is less affected from smoothening effect, $CD$ also provides a good fit.

Five sources in our sample are clearly resolved and reported earlier in literature. The fitted angular diameters are in good agreement with earlier measurements. From our analysis on light curves, we expect two more sources to be resolved (Table.\ref{ud_resolved}). One of these sources (UZ Ari) has been observed twice. There are no earlier reports of angular resolution on these two sources. The rest of the sources in our sample are unresolved. Most of them are expected to have UD values $<$4 mas.

\section*{Acknowledgments}
This work was supported by Dept of Space, Govt of India. This research made use of the SIMBAD data base operated at the CDS, Starsbourg, France and catalogs associated with it. We thank the Referee for valuable suggestions to improve the quality of the paper.

\label{lastpage}
\end{document}